\documentclass[fleqn,10pt]{wlscirep}
\usepackage[utf8]{inputenc}
\usepackage[T1]{fontenc}
\usepackage{epsfig}
\usepackage{epstopdf}
\usepackage{nicefrac}
\usepackage{graphics}
\usepackage{float}
\usepackage{amssymb}
\usepackage{graphicx,subcaption}
\usepackage{epstopdf}
\usepackage{wrapfig}
\usepackage{mathtools}
\usepackage{xcolor}
\RequirePackage{tcolorbox}
\RequirePackage[outline]{contour}
\usepackage{amsfonts}
\usepackage{dsfont}
\usepackage{bm}
\usepackage{physics}
\usepackage{verbatim}

\graphicspath{{Figure/}}
\usepackage{float}
\usepackage{graphicx,subcaption}
\usepackage{xcolor}
\RequirePackage{tcolorbox}
\RequirePackage[outline]{contour}

\renewcommand{\bra}[1]{\left\langle #1\right|}
\renewcommand{\ket}[1]{\left| #1\right\rangle}
\renewcommand{\ip}[2]{\left\langle #1 | #2\right\rangle}

\renewcommand{\tr}{\text{Tr}}
\newcommand{\tot}[1]{#1_{\text{tot}}}

\usepackage{empheq}

\definecolor{fore}{RGB}{249,242,215}
\definecolor{myblue}{rgb}{.8, .8, 1}
\definecolor{forshading}{RGB}{185,217,255}
\newcommand*{\boxedcolor}{blue}
\renewcommand{\boxed}[1]{\textcolor{\boxedcolor}{\fbox{\normalcolor\m@th$\displaystyle#1$}}}
\makeatother

\title{The role of initial system-environment correlations in the accuracies of parameters within spin-spin model}

\author[1,*]{Ali Raza Mirza}
\author[1]{Jim Al-Khalili}
\affil[1]{Department of Physics, University of Surrey, GU2 7XH, Guildford, United Kingdom}
\affil[*]{a.r.mirza@surrey.ac.uk}

\begin{abstract}
We investigate the effect of initial system-environment correlations to improve the estimation of environment parameters. By employing various physical situations of interest, we present results for the environment temperature and system-environment coupling strength. We consider the spin-spin model whereby a probe (a small controllable quantum system) interacts with a bath of quantum spins and attains a thermal equilibrium state. A projective measurement is then performed to prepare the initial state and allow it to evolve unitarily. The properties of the environment are imprinted upon the dynamics of the probe. The reduced density matrix of the probe state contains a modified decoherence factor and dissipation. This additional factor acts in such a way to improve the estimation of the environment parameters, as quantified by the quantum Fisher information (QFI). In the temperature estimation case, our results are promising as one can improve the precision of the estimates by orders of magnitude by incorporating the effect of initial correlations. The precision increases in the strong coupling regime even if the nearest neighbours' interaction is taken into account. In the case of coupling strength, interestingly the accuracy was found to be continuously increasing in both with and without correlations cases. More importantly, one can see the noticeable role of correlations in improving precision, especially at low temperatures.
\end{abstract}
\begin{document}
\flushbottom
\maketitle

\section*{Introduction}\label{intro}

Open quantum systems have garnered immense attention due to their fundamental role in the advancement of modern quantum technologies \cite{haroche2014exploring}. To properly understand quantum dynamics, it is important to scrutinize the impact of the environment, as every quantum system interacts with its surrounding environment in some way, leading to decoherence \cite{schlosshauer2007decoherence, breuer2002theory}. To comprehend this effect, it is essential to learn about the environmental parameters such as environment temperature and system-environment coupling strength. A handy approach involves using a quantum probe undergoing pure dephasing \cite{benedetti2014quantum, mirza2024improving, correa2015individual, elliott2016nondestructive,norris2016qubit,tamascelli2016characterization,streif2016measuring,benedetti2018quantum,cosco2017momentum,sone2017exact,salari2019quantum,razavian2019quantum,gebbia2020two,wu2020quantum,tamascelli2020quantum,gianani2020discrimination,mirza2023improving}. The quantum probe is allowed to interact with its surrounding environment until they both reach a correlated equilibrium state. \cite{mirza2021master}. Subsequently, a measurement is performed on the probe to prepare the desired initial state and then allow it to evolve unitarily. Studying the dynamics of such a system is challenging due to the environment's large degrees of freedom. One possible method to address this is by using exactly solvable models \cite{morozov2012decoherence, mirza2024role}. Once the dynamics are obtained, a measurement is performed on the probe that enables the estimation of various properties associated with the environment. Theoretically, a very useful parameter estimation tool is to derive quantum Fisher information, which quantifies the ultimate precision in our measurements \cite{genoni2011optical,spagnolo2012phase,pinel2013quantum,chaudhry2014utilizing,chaudhry2015detecting}. To minimize errors in the estimates, one has to maximize the Fisher information as much as possible, as dictated by the quantum Cramér-Rao bound. 

To date, many efforts have been made to estimate environmental parameters. This goal is typically achieved by initially setting the system and environment in a product state. Recent work, such as in Refs \cite{wu2020quantum, tamascelli2020quantum}, demonstrates that the environment remains in a thermal equilibrium state all the time, and the quantum correlations developed after state preparation are used to extract information about the environment. Single-qubit and two-qubit quantum probes have been employed to estimate the cutoff frequency of the harmonic oscillator environment and the spectral characterization of classical noisy environments \cite{benedetti2018quantum, benedetti2014quantum, benedetti2014characterization, mirza2024improving}. This method employs quantum Fisher information and disregards the quantum correlations in the initial system-environment equilibrium state. Improvements in the joint estimation of nonlinear coupling and nonlinearity order have been reported by squeezing quantum probes connected to nonlinear media \cite{candeloro2021quantum}. Conversely, quantum metrology employs quantum resources to enhance the sensitivity of both single and multiple-phase estimations \cite{ciampini2016quantum}. However, these findings are debatable, specifically the strong system-environment coupling regime. The importance of initial correlations present before state preparation has often been looked over \cite{uchiyama2010role,smirne2010initial,dajka2010distance,zhang2010different,tan2011non,chaudhry2013amplification,reina2014extracting,chaudhry2013role,chaudhry2014effect,zhang2015role,chen2016effects,de2017dynamics,kitajima2017expansion,buser2017initial,majeed2019effect}. More recently, \cite{mirza2024role, zhang2024improving} looked into the effects of these correlations in the estimation using QFI approach. Taking the basic seed of this idea, we extend our study to explore the effect of initial correlations in a spin environment.

We start by working out the dynamics of a single two-level system interacting with a spin environment. We allow our system and the environment to interact until they achieve a joint thermal equilibrium state. A projective measurement is then performed on the probe to prepare the initial probe state. After that, the probe interacts with the environment. We then perform a partial trace over the environment to obtain a reduced density matrix describing the dynamics of the probe. This density matrix contains the effect of decoherence and the initial correlations. With this density matrix obtained, we derive the QFI, which is a function of the interaction time between the probe and the environment. The idea here is to pick up an interaction time such that QFI is maximum. We conclusively show that the corresponding maximum QFI can be greater than the QFI obtained disregarding initial correlations. Our findings emphasize in the strong coupling at low temperatures, initial correlations play a significant role towards improving the accuracy of our measurements. This paper is organised as follows; In the first two sections, we present the scheme of state preparations and the ensuing dynamics, their details have been put in the methods sections. In the next sec., we derive the expression of quantum Fisher information for both with and without correlations cases. Then we have the main part of our paper where the results for temperature and coupling strength estimation are presented. Finally, we conclude our results in the Discussion section.

\section*{Results}\label{results}
\subsection*{Preparation of initial state}
\begin{figure}[h]
     \centering
     \includegraphics[scale=0.5,trim={0cm 0cm 0cm 0cm}]{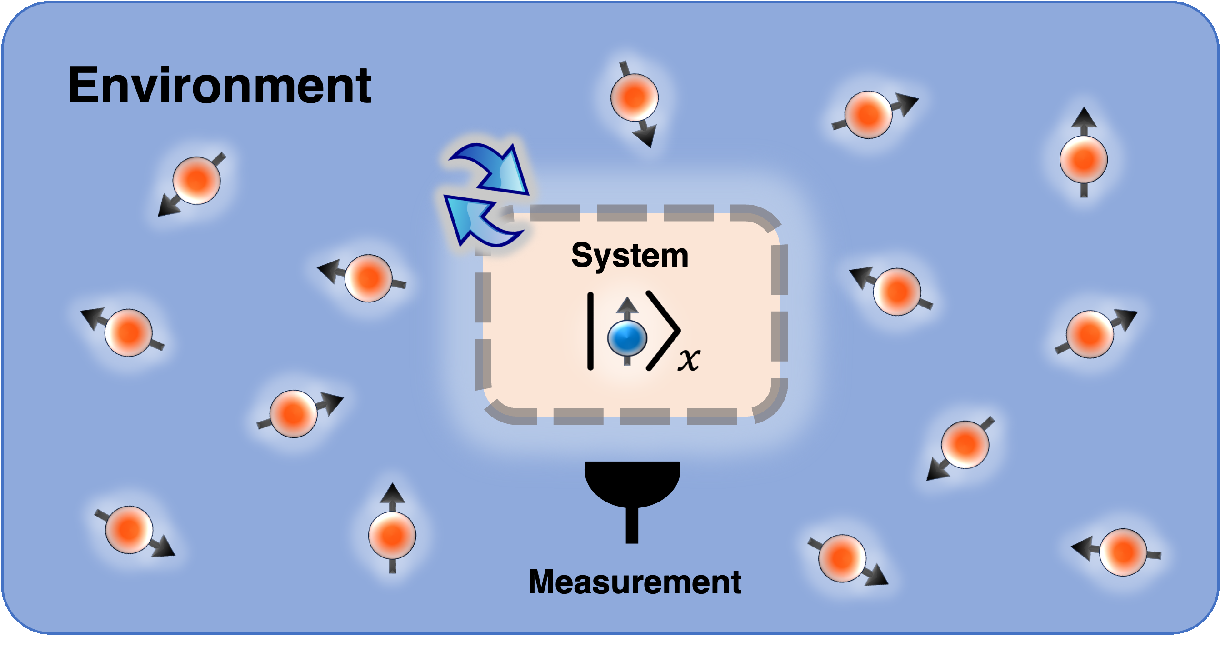}
     \caption{A model of a single two-level system (quantum probe) interacting with the collection of two-level systems}
     \label{model}
 \end{figure}

One can perceive the idea of initial probe-environment correlations by imagining that a time-independent spin system has been interacting with its surrounding environment for a very long time. The thermal equilibrium state of the system and bath is the standard canonical Gibbs state $\rho_{\text{th}}(0) = e^{-\beta \tot{H}} $, that is, we take the system–environment correlations (before the system state preparation) into account. The presence of system-environment interaction means that this state cannot, in general, be written as a product state of the system and the environment. At time $t = 0$, we then perform a projective measurement to prepare a desired system state, meaning that the initial system–environment state is now

\begin{align}
    \tot{\rho}(0)
    =\ket{\psi}\bra{\psi} \otimes \frac{ \bra{\psi} e^{-\beta \tot{H}} \ket{\psi}}{\tot{Z}}
\end{align}
with $\tot{Z} = \tr_{SE}  \left[ e^{-\beta \tot{H}} \right]$ is the partition function of the system and the environment as a whole. This is now a product state of the system and the environment. Note that the initial environment state depends on the $H$ interaction term as $SE$ as well as the initial state preparation of the system and is thus not the canonical equilibrium state for the environment. It is obvious that if the system–environment coupling is small, the initial system–environment state (after the system state preparation) would be the same as that in equation (12). In other words, for weak system–environment coupling, the effect of the initial correlations is negligible. Furthermore, the state preparation influences the initial environment state due to the initial correlations, which means that the effect of the initial correlations also depends on the system state prepared. However, the usual choice of initial state is $\tot{\rho}(0) =\ket{\psi}\bra{\psi} \otimes { e^{-\beta {H}_E} }/{Z_E}$. Here, $Z_E = \tr_{E}  \left[ e^{-\beta H_E} \right]$ is regarded as the partition function of the environment only. We refer to this initial system–environment state as the ‘uncorrelated initial state’ since the system–environment interaction before the state preparation is neglected. It is obvious that if the system–environment coupling is small, the initial system–environment state (after the system state preparation) would be the same as that in equation (12). In other words, for weak system–environment coupling, the effect of the initial correlations is negligible. Furthermore, the state preparation influences the initial environment state due to the initial correlations, which means that the effect of the initial correlations also depends on the system state prepared. In this paper, we set our projectors such that the system's initial state is  \emph{up} along \emph{x}-axis shown in Fig. \ref{model}, with or without initial correlations. Our choice of initial state corresponds to the Bloch vectors to be $\textbf{p}_x(0) =1, \textbf{p}_y(0) = 0, \textbf{p}_z(0) = 0$ for the case of correlations and ${p}_x(0) =1, {p}_y(0) = 0, {p}_z(0) = 0$ for the case of without initial correlations.

\subsection*{The model and its dynamics}
Our objective is to estimate various parameters, such as temperature and cutoff frequency, that characterize the environment of a quantum system. To achieve this, we use a two-level system coupled to the environment as a probe. By analyzing the dynamics of this probe, we can estimate the parameters characterizing the environment. Specifically, we consider a single two-level system interacting with N two-level systems. The total Hamiltonian can be written as 
    $H
    = H_0 +H_{SE},$
with $H_0 = H_{S} + H_{E}$. According to the spin-spin model
\begin{subequations}
\begin{eqnarray}
    H_S
    &=& \frac{\epsilon}{2} \sigma_z +\frac{\Delta}{2} \sigma_x,
\\
    H_{E}
    &=&\sum_{i=1}^{N} \left( \frac{\omega_{i}}{2} \sigma^{(i)}_z 
    + \chi_i \sigma^{(i)}_z \sigma^{(i+1)}_z \right),
\\
    H_{SE}
    &=& \frac{1}{2} \sigma_z \otimes g \sum_{i=1}^{N} \sigma^{(i)}_z ,
\end{eqnarray}\label{hamil}
\end{subequations}

here $\sigma_{r}$  $ (r = x, y, z) $ represent the usual Pauli spin matrices. $\epsilon$ and $\Delta$ denote the energy level spacing and the tunnelling amplitude of the central two-level system respectively. Similarly, $\omega_{i}$ denotes the energy level
spacing for the \textit{i}th environmental spin. We have also allowed the environment spins to interact with each other via $\sum_{i=1}^{N} \sigma^{(i)}_z \sigma^{(i+1)}_z \chi_i$ where $\chi_i$ characterizes the nearest neighbour interaction strength between the environment spins. The central spin interacts with the environment spins through $H_{SE}$, where $g$ is the interaction strength (assumed to be the same) between the central spin and any of the environment spins. We have also set $\hbar=1$ throughout this paper. We write $H_{SE}=S\otimes E$, where $S$ is the system operator and $E$ is the bath operator. Also we take $\ket{n}=\ket{n_1}\ket{n_2}\ket{n_3}...\ket{n_N} $ the eigenstates of $E$ with $n_i=0,1$ where $\ket{0}$ denotes spin-up state and $\ket{1}$ denotes spin-down state. Also, $[H_{E}, E]=0$, thus we have

\begin{equation}\label{eigen}
    g\sum_{i=1}^{N} \sigma^{(i)}_z\ket{n}
    = \widetilde{e}_{n} \ket{n}; 
\quad\quad\quad\quad
    \sum_{i=1}^{N} \omega_{i} \sigma^{(i)}_z\ket{n}
    = \omega_{n} \ket{n};
\quad\quad\quad\quad
    \sum_{i=1}^{N} \chi_i\sigma^{(i)}_z\sigma^{(i+1)}_z  \ket{n}
    = \alpha_{n}  \ket{n},
\end{equation}

with $\omega_{n}= \sum^N_{i=1} (-1)^{n_i} \omega_{i}$ and $\alpha_{n} = \sum^N_{i=1}\chi_i (-1)^{n_i}(-1)^{n_{i+1}}$.
with $\widetilde{e}_{n} = g \sum^N_{i=1} (-1)^{n_i}$. We assume all spins are coupled to the central spin with equal strength, all parameters having tilde overhead, meaning they are dependent on system-environment coupling strength $g$. Once we have a thermal equilibrium state, and the initial state is prepared, system dynamics can be obtained by tracing over the environment degrees of freedom $\rho(t) = \tr_{E}\left[U(t)\tot{\rho}(0)U^\dagger(t)\right]$. This demands us to first evaluate the total time evolution operator $U(t)$. We refer the reader to the consult method section for the details regarding dynamics. Now we use identity $\varrho_{c}(t) = \frac{1}{2}\left[\mathds{1} + \textbf{p}_x(t)\sigma_x + \textbf{p}_y(t)\sigma_y + \textbf{p}_z(t)\sigma_z\right]$, and after some mathematical manipulations, one can arrive the final expression of reduced density matrix
\begin{align}
    \varrho_{c}(t)
    =\frac{1}{2}  
    \left( {\begin{array}{cc}
    1 + \textbf{p}_z(t) & e^{-{\Gamma}_{c} (t)} e^{-i{\Omega_c}(t)} 
\\
    e^{-{\Gamma}_{c}(t)} e^{i{\Omega_c}(t)} & 1 - \textbf{p}_z(t)
    \end{array} } \right),\label{evo}
\end{align}
where ${\Omega_c}(t) = \arctan\left[\frac{\textbf{p}_y(t) }{\textbf{p}_x (t)} \right]$. Eigenvalues are
$\rho_{1}(t)
    = \frac{1}{2} \left[1 + \mathcal{F}_c(t) \right], $
    $ \rho_{2}(t)
    = \frac{1}{2} \left[1 - \mathcal{F}_c(t) \right],$
with
    $\mathcal{F}_c(t)
    =\sqrt{\textbf{p}^{2}_{x}(t) 
    + {\textbf{p}}^{2}_{y}(t)
    + {\textbf{p}}^{2}_{z}(t)}$. 
The corresponding eigenvectors are
\begin{subequations}
\begin{eqnarray}
    \ket{\epsilon_1(t)}
    =\sqrt{\frac{\mathcal{F}_c(t) + \textbf{p}_z(t)}{2\mathcal{F}_c(t)}} \ket{\downarrow}_z
    - e^{-i{\Omega_c} (t)} \sqrt{\frac{\mathcal{F}_c(t) - \textbf{p}_z(t)}{2\mathcal{F}_c(t)}} \ket{\uparrow}_z,
\\
    \ket{\epsilon_2(t)}
    =\sqrt{\frac{\mathcal{F}_c(t) - \textbf{p}_z(t)}{2\mathcal{F}_c(t)}} \ket{\downarrow}_z
    + e^{-i{\Omega_c} (t)} \sqrt{\frac{\mathcal{F}_c(t) + \textbf{p}_z(t)}{2\mathcal{F}_c(t)}} \ket{\uparrow}_z,
\end{eqnarray}\label{eigenvec}
\end{subequations}
where $\ket{\uparrow}_z$ and $\ket{\downarrow}_z$ are eigenstates of $\sigma_z$ with eigenvalues $+1$ and $-1$ respectively. Now solving equation $\mathcal{F}_c(t) = e^{-{\Gamma_c}}$ for $\Gamma_c$, decoherence rate comes out to be
${\Gamma}_{c}\left(t\right)
    = -\frac{1}{2}\ln \abs{\textbf{p}^{2}_{x}\left(t\right) + \textbf{p}^{2}_{y} \left(t\right)}$, where $\textbf{p}_{x}(t), \textbf{p}_{y}(t), \textbf{p}_{z}(t)$ are the usual Bloch vectors denoting the state of our qubit in the Bloch sphere representation. Their derivation is presented in the method section. If we disregard initial correlations, we obtain an analogous expression of system density matrix $\varrho_{u}(t)$ given by Eq. \eqref{dens1}. For associated analytical details, the reader is referred to the ``\hyperlink{\ref{meth}}{method}'' section.

\subsection*{Quantum Fisher Information}\label{fisher}
To quantify the precision with which a general environment parameter $\textit{x}$ can be estimated, for this we use quantum Fisher information given by \cite{benedetti2018quantum}
\begin{align}
    \mathds{F}_{q}\left(x\right)
    &=\sum_{n=1}^2 \frac{(\partial_x\rho_n)^2}{\rho_n}+2\sum_{n\neq m }\frac{(\rho_n-\rho_m)^2}{\rho_n + \rho_m}\abs{\ip{\varepsilon_m}{\partial_x \varepsilon_n}}^2,\label{genfish}
\end{align}
where $\partial_{x}$ denotes partial derivative with respect to variable \emph{x}, in our case it is temperature $T$ and system-environment coupling strength $g$. Since we have diagonalized the $\rho_c(t)$, therefore, after some algebraic manipulations, we arrive at
\begin{align}
    \mathds{F}^{c}_{q}\left(x\right)
    &=\frac{\left(\partial_x  {\Gamma_c} - \textbf{p}_z \partial_x \textbf{p}_z e^{2  {\Gamma_c}}\right)^2}{\textbf{f} \left(e^{2  {\Gamma}_c} - \textbf{f}\right)}
    + \frac{ e^{2  {\Gamma_c}} \left(\partial_x \textbf{p}_z - \textbf{p}_z \partial_x  {\Gamma_c}  \right)^2}{\textbf{f}}
    + \frac{\left( \partial_x \Omega_c \right)^2}{e^{2  {\Gamma_c}}}, \label{corrfish}
\end{align}

with $\textbf{f}
    =1 + \textbf{p}^{2}_z e^{2  {\Gamma}_c}.$ The detailed calculation of the systems' density matrix is presented in the method section \ref{corrdyn}.
Analogous expression of quantum Fisher Information disregarding the initial correlations can be written
\begin{align}
    \mathds{F}^{u}_{q}\left(x\right)
    &=\frac{\left(\partial_x  {\Gamma_u} - p_z \partial_x p_z e^{2  {\Gamma_u}}\right)^2}{f \left(e^{2  {\Gamma_u}} - f\right)}
    + \frac{ e^{2  {\Gamma_u}} \left(\partial_x p_z - p_z \partial_x  {\Gamma_u}  \right)^2}{f}
    + \frac{\left( \partial_x \Omega_u \right)^2}{e^{2 {\Gamma_u}}}, \label{unfish}
\end{align}
with $ {f} = 1 + {p}^{2}_z e^{2 {\Gamma_u}}$, ${\Omega_u}(t) = \arctan\left[\frac{{p}_y(t) }{{p}_x (t)} \right]$, and ${\Gamma_u}\left(t\right)
    = -\frac{1}{2}\ln \abs{{p}^{2}_{x}\left(t\right) 
    + {p}^{2}_{y} \left(t\right)}$. Here ${p}_{x}(t), {p}_{y}(t), {p}_{z}(t)$ are the Bloch vectors but for the case of uncorrelated initial state.

\begin{figure}[h]
    \centering
    \includegraphics[scale=1.25,trim={0cm 0cm 0cm 0cm}]{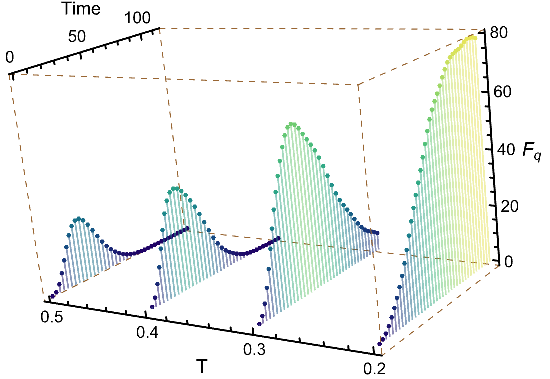}
    \caption{(Color online) Figure shows the behaviour of the quantum Fisher information $F_q$ for certain values of temperature $T$ as a function of time, incorporating the effect of the initial correlations. Peaks of $F_q$ correspond to the optimal value of the measurement estimating temperature of the environment. We have considered the environmental spins $N=50$, coupling strength $g = 0.01$, and inter-spin interaction $\chi=0$. The rest of the parameters are $\omega_i = 1, \varepsilon = 2$ and $\Delta = 1$.}
    \label{3D}
\end{figure}
\subsection*{Estimating temperature of the environment}\label{tempEst}

With the expressions of quantum Fisher information (QFI) at hand, with and without the initial correlations, we now move to estimate the temperature in this section. Since QFI is related to the Cramer–Rao bound; greater the QFI, the greater our precision of the estimate. Therefore our goal here is to maximize the QFI Eq. \eqref{corrfish} over the interaction time. As a first example, we consider estimating environment temperature. All we need to do is to calculate partial derivatives with respect to temperature $T$ using Eq. \eqref{corrfish}. We demonstrate the optimization scheme by 3D graphics in Fig. \ref{3D} where the behaviour of quantum Fisher information has been plotted as a function of interaction time for certain values of temperature. We have considered the number of spins in the environment $N=50$ and spins in the environment are assumed to be at a distance from each other such that inter-spin interactions are ignored. Peak values of $F_q$ are called optimal values at certain temperatures. The height of peaks quantifies the accuracy associated with measurement outcome, which seems to be descending. This makes sense because the quantum state is very sensitive to the temperature. As the temperature is raised, the decoherence process speeds up, we start losing the benefits of quantum properties quantum sensing is one of them which we are addressing in this paper. This figure serves as an overview of how quantum Fisher information is optimized in our later results. We work in dimensionless units where have set $\hbar=k_{B}=1$ throughout this paper.

\begin{figure}[h]
    \centering
    \begin{subfigure}[l]{0.48\linewidth}        %% or \columnwidth
        \centering
        \includegraphics[width=\linewidth]{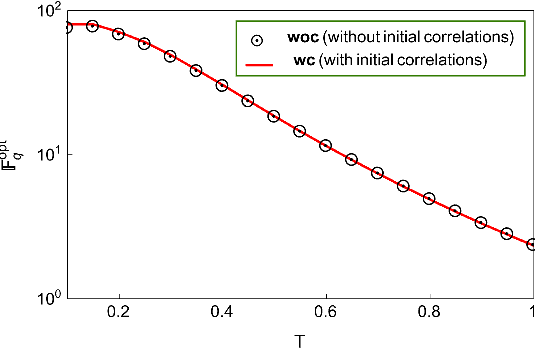}
        \caption{Coupling strength $g=0.01$}
        \label{weak}
    \end{subfigure}
    \hfill
    \begin{subfigure}[r]{0.49\linewidth}        %% or \columnwidth
        \centering
        \includegraphics[width=\linewidth]{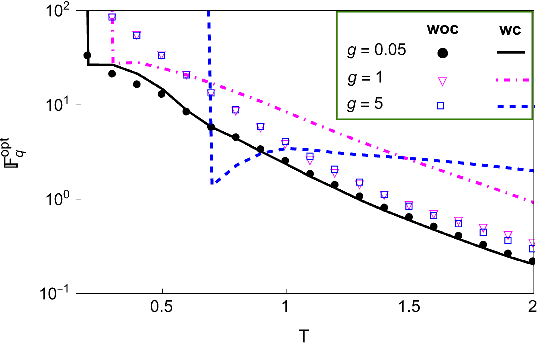}
        \caption{Coupling strength $g=0.05, 1, 5$}
        \label{coupling}
    \end{subfigure}
    \caption{(Color online) Figures show the behaviour of the optimized QFI $\mathds{F}^{\text{opt}}_{q}$ estimating the temperature of the environment with and without the effect of initial correlations. \ref{weak} considers system-environment coupling strength $g = 0.01$ while \ref{coupling} consider various  strengths $g = 0.05 (\text{black}), 1 (\text{magenta}), 5 (\text{blue})$. The rest of the system-environment parameters are the same as in Fig. \ref{3D}.}
\end{figure}

We aim to improve the precision of our estimates which can be done by plotting QFI as a function of time for each value of $T$. We do this for both cases with and without initial correlations using Eq. \eqref{corrfish} and Eq. \eqref{unfish}. The role of initial correlation is captured by the $A_n$ factor appearing in the matrix elements \eqref{mat1} given in the methods section. First, we consider system-environment coupling to be weak where the effect of correlations is minimal \cite{mirza2024role, chaudhry2013role, mirza2023improving}. This leads us to the negligible impact on the accuracy of our measurement estimates. Exactly the same trend is illustrated in Fig. \ref{weak} where we have taken $g=0.01$. We note that if we ignore the initial correlations, there is minimal effect on the precision of the measurement under consideration. The strong overlap between the curves with and without correlation is evidence. This is simply because of the contribution of the system Hamiltonian dominant as compared to the interaction Hamiltonian. However, as coupling strength increases correlation is expected to be very significant as depicted in Fig. \ref{coupling}. Here, the solid, dashed and dot-dashed curves denote QFI including correlations while curves made up of circles, squares, and triangles signify QFI disregarding correlations. At least two comments can be made regarding this result. First, at intermediate coupling strength $g=0.05$, a slight improvement in the precision can be realised for the smaller values of temperatures and then stays scenario noticeably changes as we jump into the strong coupling regime where curves belonging to the correlations are elevated than those without correlations. Therefore, it is interesting to note that precision can be significantly improved via initial correlations. as  Initial correlations play a very crucial role in the case of strong coupling.

\begin{figure}[h]
    \centering
    \begin{subfigure}[l]{0.49\linewidth}        %% or \columnwidth
        \centering
        \includegraphics[width=\linewidth]{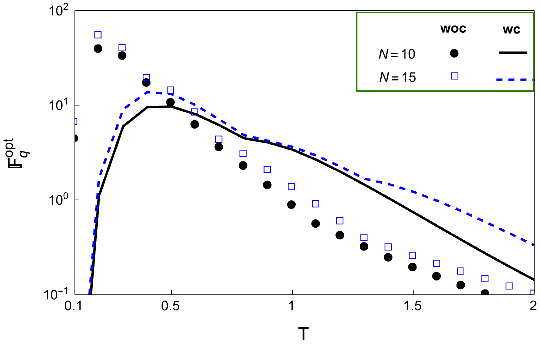}
        \caption{Effect of no. of environmental spins}
        \label{env}
    \end{subfigure}
    \hfill
    \begin{subfigure}[r]{0.49\linewidth}        %% or \columnwidth
        \centering
        \includegraphics[width=\linewidth]{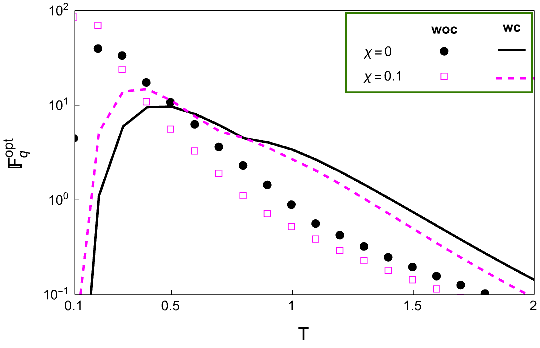}
        \caption{Effect of inter-spins interaction}
        \label{inter}
    \end{subfigure}
    \caption{(Color online) Figures illustrate the behaviour of the optimized quantum Fisher information $\mathds{F}^{\text{opt}}_{q}$ estimating the temperature of the environment with and without the effect of initial correlations. Plots in \ref{env} show the effect of the number of spins in the environment, $N=10$ Vs $N=15$, ignoring inter-spin interaction. \ref{inter} consider incorporates the effect of inter-spin interaction $\chi=0.1$ while keeping $N=10$ fixed. Both consider system-environment coupling strength $g = 0.01$ and the rest of system-environment parameters are the same as in Fig. \ref{3D}.}
\end{figure}

Next, we look into the role of the size of the environment, that is, the number of spins present in our environment. Before proceeding, we generate a standard plot for $N=10$ both with (solid-black) and without (black-dotted) initial correlations to show better comparison with $N=15$ (Fig. \ref{env}) and with $\chi=0.1$ (Fig. \ref{inter}). Since a large environment speeds up the decoherence process, thus we expect a decrease in QFI as the number of spins in the environment increases. Despite that, the results shown in Fig. \ref{env} are beyond our perception. Curves for $N=15$ seem to be lifted upwards which means a reduction in the error in our estimates. indicating Result   Similar results are obtained if we take inter-spin interaction into account. It is important to note that by including QFI slightly suppresses with or without correlation. Finally, there is something common between both Fig. \ref{env} and Fig. \ref{inter}, that is the existence of a cross over around $T=0.5$. Before this point, QFI without initial correlations either with $N=10$ or $N=15$ is greater than that belonging to correlations. However, after the cross-over point, one can witness the converse effect. Thus, it depends on the nature of the experiment being performed and on how much accuracy one would require.

\subsection*{Estimating system-environment coupling strength}\label{CoupEst}

\begin{figure}[h]
    \centering
    \begin{subfigure}[l]{0.49\linewidth}        %% or \columnwidth
        \centering
        \includegraphics[width=\linewidth]{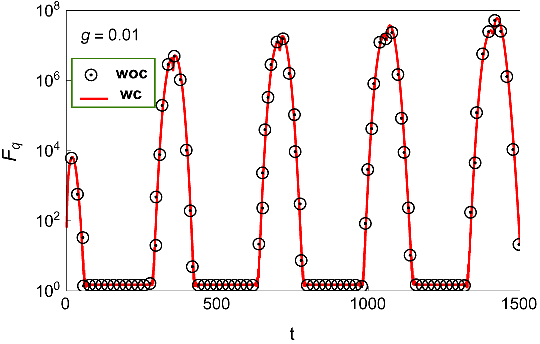}
        \caption{Temperature $T=5$, Coupling strength $g=0.01$}
        \label{high1}
    \end{subfigure}
    \hfill
    \begin{subfigure}[r]{0.49\linewidth}        %% or \columnwidth
        \centering
        \includegraphics[width=\linewidth]{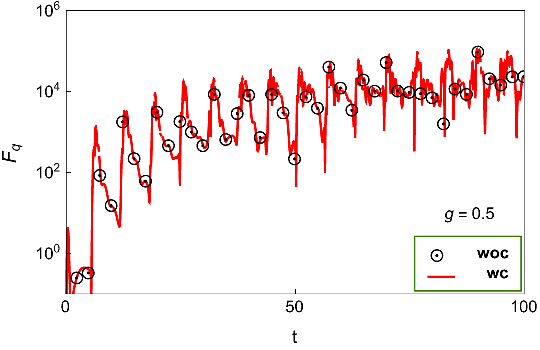}
        \caption{Temperature $T=5$, Coupling strength $g=0.5$}
        \label{high2}
    \end{subfigure}
\caption{(Color online) Figures illustrate the behavior of the QFI estimating coupling strengths $g = 0.01$ (right-hand plots) and $g = 0.5$ (right-hand plots) with and without initial correlations. The rest of parameters are the same as in Fig. \ref{3D}. }\label{high}
\end{figure}

\begin{figure}[h]
    \centering
    \begin{subfigure}[l]{0.49\linewidth}        %% or \columnwidth
        \centering
        \includegraphics[width=\linewidth]{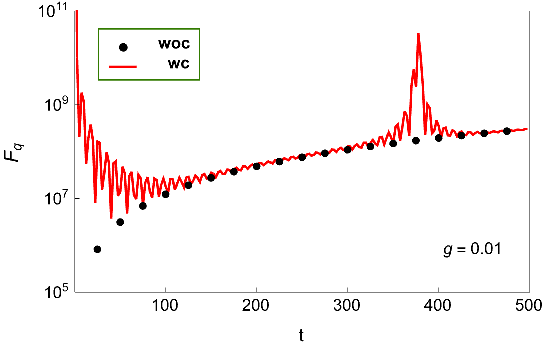}
        % \caption{$T=0.1$ Coupling strength $g=0.01$}
        % \label{jg}
    \end{subfigure}
\hfill
    \begin{subfigure}[r]{0.49\linewidth}        %% or \columnwidth
        \centering
        \includegraphics[width=\linewidth]{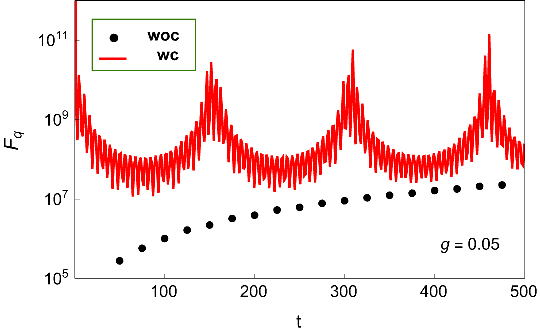}
        % \caption{$T=0.1$ Coupling strength $g=0.05$}
    \end{subfigure}
    \caption{(Color online) Same as Fig. \ref{high} except that now we have considered low temperature $T=0.1$ for both of above.}
    \label{low}
\end{figure}

Next, we consider estimating coupling strength. Once again we use the same expression given in Eq. \eqref{corrfish} and \eqref{unfish}. Here we need derivatives w.r.t. coupling strength $g$. and we optimize it over the interaction time to get Optimized Quantum Fisher Information. We compare QFI for estimating the coupling strength $g$ obtained without initial correlation with the QFI obtained with initial correlation, at $T$ high temperature. Results are illustrated in Fig. \ref{high1}, where we have shown the QFI as a function of interaction time, setting coupling strength to be fixed at $g=0.01$. (Color Online) Red-solid curve denotes the results including correlations (wc) while black-dotted circles signify the $F_q$ disregarding correlations (woc). At least two points should be noted here. First, unlike the case of temperature estimation, here $F_{q}$ keeps on increasing with time. It means interaction time has to be prolonged by the required precision. Second, The overlap of \emph{woc}, \emph{wc} curves showing correlations are not a matter of any benefit. 
A similar trend remains in the stronger coupling Fig. \ref{high2}, where can still see the overlaps. In both of these cases, we can witness no appreciable quantitative difference between both with or without correlations. However, if we compare Fig. \ref{high1} and Fig. \ref{high2} with each other, we can see decrement of QFI by the order of magnitude at stronger coupling $g=0.5$. This is because strong coupling speeds up the decoherence process at higher temperatures, and our QFI is directly linked with the decoherence functions $\Gamma_u$ and $\Gamma_c$ in Eq. \eqref{unfish} and Eq. \eqref{corrfish} respectively.
Next, we move on to the low temperatures regime, where correlation time is longer in comparison with at high temperatures. The reason is obvious that the sensitive quantum phase is no longer exposed to the thermal energies, hence quantumness can be preserved. Therefore, we expect a noticeable improvement in the quantum measurement if initial correlations are taken into account. Our forecast is illustrated in Fig. \ref{low}. We start from the weak coupling $g=0.01$, one thing is again common, the continuous increase of QFI with time, however, we can now better differentiate \emph{woc}, \emph{wc} and curves. This difference is further amplified if we slightly tune up the coupling to $g=0.05$. This difference entails the improvement in accuracy which is the ultimate goal.

\section*{Discussion}\label{discs}

Based on the estimation results presented in this paper, we deduce the effect of initial system-environment correlations is somehow advantageous in the estimation of environment parameters that is the environment's temperature and system-environment coupling strength. In the temperature estimation, the precision does not significantly increase with the increase of coupling strength, but we noticeable improvement can be seen in the case of initial correlations. As we reduce the number of spins in the environment, QFI also seems to be decreasing in both cases under consideration. A similar trend has been seen if the inter-spin interaction is taken into account. As the quantitative difference is not appreciable, thus ignoring the nearest neighbours' interaction is a reasonable assumption within the chosen set of system-environment parameters. On the other hand, in the case of coupling strength estimation, our results are promising as one can see that $F_{q}$ is found to be continuously increasing in both with and without correlations cases. By incorporating the effect of initial correlations, one can witness that the precision of the estimates increases by orders of magnitude, especially at low temperatures.

\section*{Methods}\label{meth}
\subsection*{System dynamics without initial correlations}\label{undyn}

Here we assume that the system and the environment are initially in the product state $\rho = \rho_S \otimes \rho_E $ whereby no quantum correlations are present. Now we prepare our system initial state \emph{up} along \emph{x}-axis via projective measurement. Now evolution $\rho(t) = \tr_{E}\left[U(t)\tot{\rho}(0)U^\dagger(t)\right]$ without initial correlations, we have reduced density matrix $\rho_{u}(t)$
\begin{align}
    \rho_{u}(t)
    &=\frac{1}{Z_E}\sum_{n}  c_{n} U_{n}(t)\ket{\psi}\bra{\psi}  U^\dagger_{n}(t).
\end{align}
It is convenient to write this state in terms of Bloch vector components in matrix form as
\begin{equation}
    \left( {\begin{array}{c}
    {p}_{x} (t)\\
    {p}_{y} (t)\\
    {p}_{z} (t)
    \end{array} } \right) = \frac{1}{Z_E} 
    \left( {\begin{array}{ccc}
    {M}_{xx} & {M}_{xy} & {M}_{xz} 
\\
    {M}_{yx} & {M}_{yy} & {M}_{yz} 
\\
    {M}_{zx} & {M}_{zy} & {M}_{zz}
    \end{array} } \right) 
    \left( {\begin{array}{c}
    p_x(0)\\
    p_y(0)\\
    p_z(0)
  \end{array} } \right),
\end{equation}
with
\begin{equation*}
  \begin{split}
    {M}_{xx} 
    &=\sum_{n}\frac{ c_{n}}{4{\eta}^2_{n}}\Big\{\Delta^2 + {\eta}_{n}^{2}\cos(2{\eta}_{n}t)\Big\};
\\
    {M}_{yx}  
    &= \sum_{n} \frac{c_{n} \varepsilon_n}{2{\eta}_{n}}  \sin(2{\eta}_{n} t);
\\
    {M}_{zx} 
    &=\sum_{n} \frac{c_{n} \widetilde{\epsilon}_{n}\Delta}{2{\eta}^2_{n}}\sin^2\left({\eta}_{n}t\right);
  \end{split}
\quad\quad\quad
  \begin{split}
    {M}_{xy} 
    &= -\sum_{n} \frac{ c_{n} \widetilde{\epsilon}_{n}}{2{\eta}_{n}}\sin\left(2{\eta}_{n}t\right);
\\
    {M}_{yy}  
    &= \sum_{n} c_{n} \cos(2{\eta}_{n} t);
\\
    {M}_{zy} 
    &= \sum_{n} \frac{ c_{n} \Delta}{2{\eta}_{n}} \sin(2{\eta}_{n} t);
  \end{split}
\quad\quad\quad
  \begin{split}
    {M}_{xz} 
    &=\sum_{n} \frac{ c_{n} \widetilde{\epsilon}_{n}\Delta}{2{\eta}^2_{n}}\sin^2\left({\eta}_{n}t\right);
\\
    {M}_{yz} 
    &= - \sum_{n} \frac{c_{n} \Delta}{2{\eta}_{n}}\sin\left(2{\eta}_{n}t\right);
\\
    {M}_{zz} 
    &=\sum_{n}\frac{c_{n}}{4{\eta}^2_{n}}\Big\{{\eta}_{n}^{2}+\Delta^2\cos(2{\eta}_{n}t)\Big\}.
  \end{split}\label{ele2}
\end{equation*}
here we have $Z_{E}=\sum_{n}{c_{n}}$. Thus the reduced density matrix is given by
\begin{align}
    \varrho_{u}(t)
    =\frac{1}{2}  
    \left( {\begin{array}{cc}
    1 + {p}_z(t) & e^{-{\Gamma}_{u} (t)} e^{-i{\Omega}_{u}(t)} 
\\
    e^{-{\Omega}_{u}(t)} e^{i{\phi}(t)} & 1 - {p}_z(t)
    \end{array} } \right).\label{dens1}
\end{align}
Eigenvalues are
$\rho_{1}(t)
    = \frac{1}{2} \left[1 + \mathcal{F}_u(t) \right], $ 
    $ \rho_{2}(t)
    = \frac{1}{2} \left[1 - \mathcal{F}_u(t) \right],$
with
    $\mathcal{F}_u(t)
    =\sqrt{ {p}^{2}_{x}(t) 
    + { {p}}^{2}_{y}(t)
    + { {p}}^{2}_{z}(t)}$. 
The corresponding eigenvectors are
\begin{subequations}
\begin{eqnarray}
    \ket{\epsilon_1(t)}
    =\sqrt{\frac{\mathcal{F}_u(t) +  {p}_z(t)}{2\mathcal{F}_u(t)}} \ket{\downarrow}_z
    - e^{-i{\Omega_u} (t)} \sqrt{\frac{\mathcal{F}_u(t) -  {p}_z(t)}{2\mathcal{F}_u(t)}} \ket{\uparrow}_z,
\\
    \ket{\epsilon_2(t)}
    =\sqrt{\frac{\mathcal{F}_u(t) -  {p}_z(t)}{2\mathcal{F}_u(t)}} \ket{\downarrow}_z
    + e^{-i{\Omega_u} (t)} \sqrt{\frac{\mathcal{F}_u(t) +  {p}_z(t)}{2\mathcal{F}_u(t)}} \ket{\uparrow}_z,
\end{eqnarray}\label{eigenvec}
\end{subequations}

\subsection*{System dynamics with initial correlations}\label{corrdyn}

One imagines that a time-independent spin system has been interacting with its surrounding environment for a very long time. The thermal equilibrium state of the system and bath is the standard canonical Gibbs state $\tot{\rho}(0)$. For such a state, system dynamics can be obtained by tracing over the environment degrees of freedom $\rho_{c}(t) = \tr_{E}\left[U(t)\tot{\rho}(0)U^\dagger(t)\right]$. This demands to evaluate first the total time evolution operator $U(t)$ as
\begin{align}
    U(t)
    =\sum_{n} e^{-i\frac{\omega_{n}}2t}e^{-i\alpha_{n} t}e^{-i\widetilde{H}^{n}_{s}t}\ket{n}\bra{n},
    =\sum_{n} U_{n}(t)\ket{n}\bra{n},\label{unitary}
\end{align}
with $U_{n}(t) = e^{-i\frac{\omega_{n}}2t}e^{-i\alpha_{n} t}e^{-i\widetilde{H}^{n}_{s}t}$ and $\widetilde{H}^{n}_{s}
    \equiv \frac{\widetilde{\epsilon}_{n}}{2} \sigma_z +\frac{\Delta}{2} \sigma_x,$ as a shifted Hamiltonian, with $\widetilde{\epsilon}_{n} = \epsilon + g \sum^N_{i=1} (-1)^{n_i} \equiv \epsilon + \widetilde{e}_{n} .$ Now using identity \footnote{$e^{ia(\hat{n}\cdot\Vec{\sigma})} = \mathds{1}\cos a + i(\hat{n}\cdot\Vec{\sigma})\sin a$} to expand $U_{n}(t)$ as
    $U_{n}(t)
    =e^{-i\frac{\omega_{n}}2t}e^{-i\alpha_{n} t} \left\{\cos({\eta}_{n}t)
    - \frac{i\sin({\eta}_{n}t)}{{\eta}_{n}} \widetilde{H}^{n}_{s} \right\},$ only acting on the systems Hilbert space. Also, we have set ${\eta}_{n} = \frac{1}{2}\sqrt{\widetilde{\epsilon}_{n}^2 + \Delta^2} $. 
where $c_{n}=e^{-\beta\frac{\omega_{n}}{2}}e^{-\beta\alpha_{n} }$, and $\tot{Z} = \sum_{n}{A_{n} c_{n}}$ with $A_{n} = \cosh(\beta \eta_{n}) - \frac{\cosh(\beta \eta_{n})}{\eta_{n}} \bra{\psi} \widetilde{H}^{n}_{s} \ket{\psi} $
\begin{align}
    \rho_{c} (t)
    =\tr_{E}\left[U(t)\tot{\rho}(0)U^\dagger(t)\right]
    = \frac{1}{\tot{Z}} \sum_{n}A_{n} c_{n} U_{n}(t) \ket{\psi}\bra{\psi}U^\dagger_{n}(t).
\end{align}
Casting the associated the Bloch vectors into matrix form as
\begin{equation}
    \left( {\begin{array}{c}
    \mathbf{p}_{x} (t)\\
    \mathbf{p}_{y} (t)\\
    \mathbf{p}_{z} (t)
    \end{array} } \right) = \frac{1}{\tot{Z}} 
    \left( {\begin{array}{ccc}
    \mathbf{M}_{xx} & \mathbf{M}_{xy} & \mathbf{M}_{xz} 
\\
    \mathbf{M}_{yx} & \mathbf{M}_{yy} & \mathbf{M}_{yz} 
\\
    \mathbf{M}_{zx} & \mathbf{M}_{zy} & \mathbf{M}_{zz}
    \end{array} } \right) 
    \left( {\begin{array}{c}
    \mathbf{p}_x(0)\\
    \mathbf{p}_y(0)\\
    \mathbf{p}_z(0)
  \end{array} } \right),
\end{equation}
which can now be written in terms of density matrix given in the main text Eq. \eqref{evo}. Matrix elements are;
\begin{equation}\label{mat1}
  \begin{split}
    \mathbf{M}_{xx} 
    &=\sum_{n}\frac{A_{n} c_{n}}{4{\eta}^2_{n}}\Big\{\Delta^2 + {\eta}_{n}^{2}\cos(2{\eta}_{n}t)\Big\};
\\
    \mathbf{M}_{yx} 
    &=\sum_{n} \frac{A_{n} c_{n} \widetilde{\epsilon}_{n}}{2{\eta}_{n}}\sin\left(2{\eta}_{n}t\right);
\\
    \mathbf{M}_{zx} 
    &=\sum_{n} \frac{A_{n} c_{n} \widetilde{\epsilon}_{n}\Delta}{2{\eta}^2_{n}}\sin^2\left({\eta}_{n}t\right);
  \end{split}
\quad\quad\quad
  \begin{split}
    \mathbf{M}_{xy}  
    &= -\sum_{n} \frac{A_{n} c_{n} \varepsilon_n}{2{\eta}_{n}}  \sin(2{\eta}_{n} t);
\\
    \mathbf{M}_{yy}  
    &= \sum_{n} A_{n} c_{n} \cos(2{\eta}_{n} t);
\\
    \mathbf{M}_{zy} 
    &= \sum_{n} \frac{A_{n} c_{n} \Delta}{2{\eta}_{n}} \sin(2{\eta}_{n} t);
  \end{split}
\quad\quad\quad
  \begin{split}
    \mathbf{M}_{xz} 
    &=\sum_{n} \frac{A_{n} c_{n} \widetilde{\epsilon}_{n}\Delta}{2{\eta}^2_{n}}\sin^2\left({\eta}_{n}t\right);
\\
    \mathbf{M}_{yz} 
    &= - \sum_{n} \frac{A_{n} c_{n} \Delta}{2{\eta}_{n}}\sin\left(2{\eta}_{n}t\right);
\\
    \mathbf{M}_{zz} 
    &=\sum_{n}\frac{A_{n} c_{n}}{4{\eta}^2_{n}}\Big\{{\eta}_{n}^{2}+\Delta^2\cos(2{\eta}_{n}t)\Big\}.
  \end{split}
\end{equation}

\bibliography{main}

\section*{Acknowledgement}

A.~R.~Mirza and J. Al-Khalili are grateful for support from the John Templeton Foundation Trust. We also acknowledge useful discussions with Dr Adam Zaman Chaudhry and Dr Ahsan Nazir.

\end{document}